\documentclass[a4paper,fleqn]{cas-dc}

\usepackage[authoryear]{natbib}

\usepackage{graphicx}	
\usepackage{amsmath}	

\usepackage{ulem}

\def\be{\begin{equation}}
\def\ee{\end{equation}}
\def\ba{\begin{eqnarray}}
\def\ea{\end{eqnarray}}

\def\msun{M_\odot}

\def\ltsima{$\; \buildrel < \over \sim \;$}
\def\simlt{\lower.5ex\hbox{\ltsima}}
\def\gtsima{$\; \buildrel > \over \sim \;$}
\def\simgt{\lower.5ex\hbox{\gtsima}}

\definecolor{webgreen}{rgb}{0,.5,0}
\definecolor{webbrown}{rgb}{.6,0,0}
\definecolor{falured}{rgb}{0.5, 0.09, 0.09}

\definecolor{darkblue}{rgb}{0.1, 0.1, 0.6}

\definecolor{darkgreen}{rgb}{0.1, 0.4, 0.0}

\begin{document}
\let\WriteBookmarks\relax
\def\floatpagepagefraction{1}
\def\textpagefraction{.001}
\shorttitle{Dust destruction in bubbles driven by multiple-SNe}
\shortauthors{E.O. Vasiliev \& B.B. Nath}

\title [mode = title]{Dust destruction in bubbles driven by multiple supernovae explosions}                      

\author[1]{Evgenii O. Vasiliev}
\cormark[1]
\ead{eugstar@mail.ru}
\affiliation[1]{organization={Lebedev Physical Institute, Russian Academy of Sciences},
                addressline={53 Leninsky Avenue}, 
                city={Moscow},
                postcode={119991}, 
                country={Russia}}

\author[2]{Biman B. Nath}
\affiliation[2]{organization={Department of Physical Sciences, Indian Institute of Science Education and Research (IISER)},
                addressline={Knowledge City, Sector 81, Sahibzada Ajit Singh Nagar}, 
                postcode={140306}, 
                city={Mohali},
                state={Punjab}, 
                country={India}}

\cortext[cor1]{Corresponding author}

\begin{abstract}
Dust lifetime derived from an isolated supernova (SN) evolution in the interstellar medium is known to be an order of magnitude shorter than the time needed to replenish dust mass by its production in various Galactic sources. We show, with the help of 3-D numerical hydrodynamical simulations, that destruction of dust in the case of multiple SNe in a star cluster is markedly different from that in an isolated SN. We find that the mass of dust destroyed in the bubble does not grow for a considerable time, while SNe continue to explode. This regime is attained at saturation timescale, which is proportional to SNe rate in cluster. We show that the mass of dust destroyed in bubble per SN decreases for higher SN rates. Thus, the relative destruction efficiency -- defined as the ratio of the the total mass of dust destroyed by clustered SNe to that destroyed by the same number of isolated SNe -- in bubbles evolved in a homogeneous medium drops for massive clusters, e.g. around clusters with $M_\ast \gtrsim 4\times 10^4 M_\odot$ it is less $0.4$\%. For lower mass clusters, the efficiency is proportional to the average time delay between SNe. We found that SNe in a cluster with $M_\ast \lesssim 4\times 10^4 M_\odot$ destroy the same amount of dust as if they were exploding in isolation. In a clumpy medium with lognormal distribution of density fluctuations having dispersion $\sigma\sim 2.2$ and a spatial Kolmogorov power-law spectrum, the efficiency increases by a factor of $\sim 1.5$ in bubbles formed around clusters with $M_\ast \sim 4\times 10^4 M_\odot$ and up to 4 times around $M_\ast \sim 8\times 10^5 M_\odot$. We argue that the interstellar dust swept up by multiple SNe almost completely survives in the shells of bubbles around such massive clusters. Therefore, the destruction of the interstellar dust is controlled by SNe in low-mass clusters. We conclude that the interstellar dust lifetime for a given SN rate is at least a factor $\sim 10$ longer as compared to the estimates derived from an isolated SN.
\end{abstract}


\begin{keywords}
interstellar dust \sep interstellar dust processes \sep supernova remnants
\end{keywords}

\maketitle

\section{Introduction} 
\label{intro}

Interstellar dust is thought to be destroyed, among the processes, by the inertial and thermal sputtering in hot gas \citep[][and references therein]{Barlow1978,Draine1979a,Draine1979b,McKee1989,Jones1994,Nath2008,Slavin2015,Priestley2022}. In cold gas, surviving dust can be shattered in grain-grain collisions at higher densities, replenishing the population of smaller grains \citep[][]{Borkowski1995,Jones1996,Slavin2004,Guillet2009,Bocchio2016,Kirchschlager2019}. During further evolution, these grains pre-existing in the interstellar medium can be entrained by strong shocks and may again be found in hot environment. The typical rate of shocks in the Galactic interstellar medium implies a characteristic dust lifetime that ranges between $3\times 10^8 - 10^9$~yr \citep[][]{McKee1989,Jones1994,Jones1994b,Slavin2015}. These estimates are based on the conclusion that each SN has 'cleaned out' $\sim 1300~\msun$ of the interstellar gas, so that it has destroyed $\sim13\msun$ of the interstellar dust for the dust-to-gas ratio of 0.01 \citep{McKee1989}. Such short time scale,  compared to the dust production time \citep{McKee1989,Draine2009}, raises an important question regarding the dust mass budget in the ISM \citep[see for recent discussion in][]{Mattsson2021,Kirchschlager2022,Peroux2023}, which is commonly referred to as the {\it `dust budget crisis'} \citep[][]{Michalowski2010,Dunne2011,Rowlands2014,Galliano2021}.

The problem can be briefly explained as follows. Dust is efficiently sputtered by shocks with speeds as high as $\simgt 200$~km/s. Such speeds are typical for SN expansion at adiabatic phase in the diffuse interstellar gas. During this period, the kinetic energy of SN shell is of the order of the SN energy: $Mv_s^2 \sim E_{SN}$. This results in an estimate of the swept-up interstellar material by a single isolated SN of about $10^3~\msun$ \citep{McKee1989,Draine2009}. From this, the destroyed dust mass can be estimated to be of order of $10~\msun$ per SN, if the gas-to-dust mass is $\sim 100$. Using this result, the authors \citep{McKee1989} obtained the dust lifetime for the Galactic SN rate as $\tau \sim 4\times 10^8$~yr. This timescale is shorter by than a factor around 20 the dust replenishment time from AGB stars and SNe \citep{Draine2009}. The mismatch leads to rapid reduction of dust amount in the interstellar medium, while the medium is rich in dust.

This estimate is, however, valid only for isolated SN explosions. In the case of multiple SNe, which is more common than isolated SN, because stellar SN progenitors are typically located in OB-associations \citep[e.g.,][]{Krumholz2019}, the situation can be different.  In the bubble produced by multiple SNe explosions,  the efficiency of  dust destruction  can be markedly different because subsequent explosions occur inside hot rarefied gas, so that when the shock fronts finally interact with a dense shell, the shocks would have lost much of their energy. A similar scenario occurs when a SN explodes in a wind-driven bubble \citep{Martinez2019}. The speed of shocks of the  subsequent SNe falls  below $200$~km/s after interacting with the wind-driven dense shell. Therefore, one can expect that the damped shocks lose their ability to destroy dust in the shell, at least in the case when the shocks are spherically symmetric.

Here we consider in detail for the first time the destruction of dust in the shells driven by multi-SNe. The paper is organized as follows. Section~\ref{sec:model} describes the model details. Section~\ref{sec:dynamics} presents the evolution of the swept-up interstellar dust in bubbles driven by multiple-SNe explosions in stellar clusters. In Section~\ref{sec:discus} we summarize our results and  possible consequences.


\section{Model description} \label{sec:model}

Let us consider a star cluster of mass $M_\ast$. The number of SN in it is $N_{SN} \sim M_\ast/150~\msun$ for a Salpeter initial mass function (IMF) for stellar masses in the range of $0.1-40~\msun$, i.e. one SN explosion per $150~\msun$ star formation. SNe progenitors are thought to have masses in range $8-40~\msun$. The duration of SNe explosions spans the lifetime of the least massive star of $8\msun$ $t_{lf,max}\sim 24$~Myr ($t_{lf} \sim M^{-1.6}$) \citep{Iben-book2}. The average time between SN explosions in a cluster is, therefore,  $\Delta t \sim 24~{\rm Myr} / N_{SN}$. For example, in a cluster of $M_\ast \sim 4\times 10^4~\msun$, the number of SNe is $N_{SN} \sim 250$ and the average time delay between explosions is $\Delta t \sim 10^5$~yr. 

The mass function of stellar clusters is $dN/dM \propto M^{-\alpha}$, where $\alpha = 1.25 - 2.25$, $M_{min}\sim 10^3~\msun$, $M_{max} \sim 3\times 10^7\msun$ \citep{Krumholz2019}. The estimate above shows that the higher the cluster mass $M_\ast$, the higher the SN rate, or, equivalently, the shorter the time delay between explosions: $\Delta t\sim 4000~{\rm Myr}/M_\ast$. Our models with different $\Delta t$ describe  clusters of varying masses, since $M_\ast \propto \Delta t^{-1}$. Equal time gaps between subsequent SNe explosions are more or less statistically justified \citep[e.g.,][]{Ferrand2010} and widely used \citep[e.g.,][]{Kim2017,Yadav2017}. The statistical modelling of stellar masses for a given stellar cluster, i.e. a sequence of SN explosions in a cluster, shows that in massive clusters the time-delays between subsequent SN explosions are almost the same during first $10-12$~Myr. In average these delays are close to each other, thus, we can assume a mean value for this delay. Note that in less massive clusters with $M\simlt 8\times 10^3~\msun$ the delay between SNe explosions becomes irregular due to rarity of massive stars.

We carry out 3D hydrodynamic simulations of the ISM under the action of multiple clustered SNe explosions. SNe are distributed uniformly and randomly inside a region with a fixed cluster radius $r_c = 10$~pc. Note that this value is close to the half-mass radius of stellar clusters $r_h$ with mass $\sim 10^7\msun$ (the $r_h - M_\ast$ relation increases slightly more slowly than $r_h \propto M_\ast^{1/3}$ \citep[see Figure~9 in][]{Krumholz2019}). Although the cluster radius varies between several to 10~pc for clusters with masses in the range of $\sim 10^4\hbox{--}10^7$ M$_\odot$, our assumption of a fixed radius does not pose a problem in our models. Note that the radius of the bubble formed by the first SN explosion reaches about 10~pc at the age of 10~kyr in a medium with density of $\sim 1$~cm$^{-3}$. Thus, subsequent SN will explode in the hot bubble and there is no difference in the bubble dynamics as long as the cluster has radius smaller 30~pc in a medium with density of $\sim 1$~cm$^{-3}$ \citep{vsn2017}.

We have assumed one SN per $\Delta t = 100$, 50, 30, 10 and 5~kyr, which correspond to  clusters of masses $M_\ast \sim 4\times 10^4, \ 8\times 10^4, \ 1.2\times 10^5, \ 4\times 10^5$ and $8\times 10^5~\msun$, the last one being the upper limit of stellar cluster mass in the Milky Way \citep[see, e.g.][]{Krumholz2019}. We inject the mass and energy of a SN in a region of radius $r_0=3$~pc, assuming commonly used values $M_{inj}=30~\msun$ and $E=10^{51}$~erg. The energy is injected in thermal form, although the results are not sensitive to the form of injected energy \citep{Sharma2014}.

We consider primarily single-sized interstellar dust with size equal to $a_0 = 0.1~\mu$m. During the evolution, dust grains are destroyed and their sizes decrease continuosly (no binning) depending on physical conditions in ambient gas; the minimum size is set to $0.001~\mu$m. {Our model with single-sized dust is motivated by the following arguments: i) 0.1~$\mu$m dust particles are  typical in the ISM \citep{Draine2003}; ii) dust particles with radii $a \simlt 0.05~\mu$m are efficiently sputtered through the action of SN shocks, and iii) such an approach considerably simplifies interpretation of the numerical results and their analysis. Moreover, the survival dust mass fraction in models with monodisperse and polydisperse \citep[MRN type,][]{Mathis1977} dust-size distributions are shown to be comparable \citep{Dedikov2025na}.}

The gas distribution in the ambient (background) medium is set to be homogeneous with density $n_b = 1$~cm$^{-3}$ as a fiducial value, the temperature in all models is $10^4$~K. We study the case of a clumpy medium in a later section 3.1. The metallicity of the background gas is equal to solar value ${\rm [Z/H]} = 0$. We assume a dust-to-gas (DtG) mass ratio equal to $\zeta_b=0.01$.

Simulations are run with tabulated non-equilibrium cooling rates for a gas that cools isochorically from $10^8$ down to 10~K \citep{v11,v13}. The heating rate is assumed to be constant, with a value chosen such as to stabilize the radiative cooling of the ambient gas at $T=10^4$~K. This diffuse heating term represents the photoelectric heating of dust grains \citep{Bakes1994}, which is thought to be the dominant heating mechanism in the interstellar medium.

We use our gas-dynamic code \citep{vns2015,vsn2017} based on the unsplit total variation diminishing (TVD) approach that provides high-resolution capturing of shocks and prevents unphysical oscillations, and the Monotonic Upstream-Centered Scheme for Conservation Laws (MUSCL)-Hancock scheme with the Haarten-Lax-van Leer-Contact (HLLC) method \citep[see e.g.][]{Toro2009} as approximate Riemann solver. This code has successfully passed the whole set of tests proposed in \citep{Klingenberg2007}. In order to follow the dynamics of dust particles we have implemented the method \citep[see description and tests in Appendix A of][]{vs2024} similar to that proposed in the papers \citep{Youdin2007,Mignone2019,Moseley2023}. Dust particles are described by several features (e.g., mass, size, sort, id number etc) to allow one to identify their evolution and physical properties. To follow the transport of dust mass (not of an individual grain) in a gaseous flow we introduce a macroscopic mass of a `superparticle' or macroparticle, which represents a conglomerate of microscopic grains. Thus, the dust component is modelled as an ensemble of macroparticles  governed by the motion equations \citep[see description in Appendix A of][]{vs2024}. The backward reaction of dust on to gas due to momentum transfer, work done by the drag force and the frictional heating from dust particles are also accounted for in order to ensure both dynamical and thermal self-consistency. We take into account the destruction of dust particles by both thermal (in a hot gas) and kinetic (due to a relative motion between gas and grains) sputtering \citep{Draine1979b}. 

The mass of dust in a medium is distributed between many dust `superparticles'. Each `superparticle' is supposed to represent an ensemble of numerous physical grains of a single size. To follow the dust transport in a medium we set at least one `superparticle'  per computational cell. Each `superparticle' contains the total mass of the ensemble of single-sized dust particles in it. So the total dust mass in a cell is a sum of masses of `superparticles' inside a cell \citep[see for more details][]{vs2024}.

The cell size equal to 1~pc is adopted for the runs with ambient gas number density $n = 1$~cm$^{-3}$. This 
is sufficient both for resolving the thermal (cooling) length estimated as $\lambda_t \sim 5~n^{-1} T_6$~pc and for reproducing the global bubble dynamics driven by multiple SNe \citep[e.g.,][]{Fielding2018}. The numerical grid consists of 256$^3$ cells. This results in the total number of interstellar dust `superparticles' in the domain being $256^3 \sim 16$ millions. 

\begin{figure}
\center
\includegraphics[width=8.5cm]{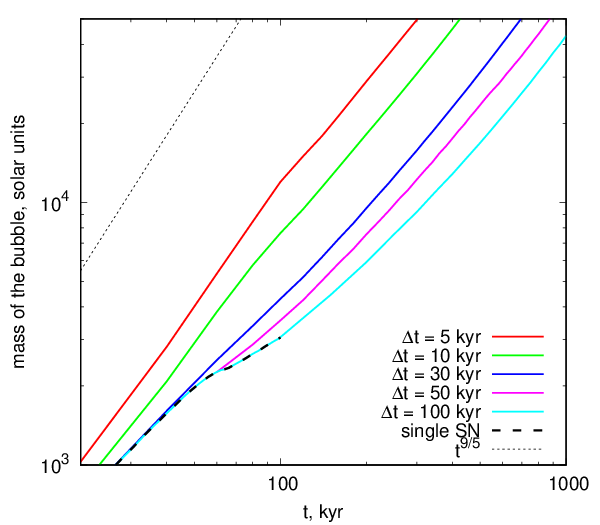}
\caption{
Mass of the bubble driven by multiple SNe explosions in a cluster. The color lines depict the evolution for different rates, of one SN per $\Delta t = 100$, 50, 30, 10 and 5~kyr, which correspond to clusters of masses $M_\ast \sim 4\times 10^4, \ 8\times 10^4, \ 1.2\times 10^5, \ 4\times 10^5$ and $8\times 10^5~\msun$. The dashed line (only until 100~kyr) shows the evolution for the bubble formed by single SN explosion.
The dotted line shows the slope of $t^{9/5}$ for comparison.
}
\label{fig-mgas-evol}
\end{figure}

\section{Dynamics} \label{sec:dynamics} 

Fig.~\ref{fig-mgas-evol} presents the time evolution of the total mass of gas within the bubble (including injected and swept up mass), created by multiple SNe exploded in a cluster with different SN rate. The dashed line (which is run for only 100 kyr) shows the evolution for the bubble formed by single SN explosion. The total bubble mass grows closely as $t^{9/5}$ (depicted by dotted line), which is expected from the self-similar evolution of wind bubbles ($r\propto t^{3/5}$). The differences from the self-similar evolution are due to efficient cooling in the bubble shell \citep[e.g.,][]{vsn2017}. The bubble driven by SNe of rate $(100{\rm kyr})^{-1}$, for example, attains this slope after $t \sim 400-500$~kyr. More massive clusters attain the wind regime at earlier epochs.  

\begin{figure}
\center
\includegraphics[width=8.5cm]{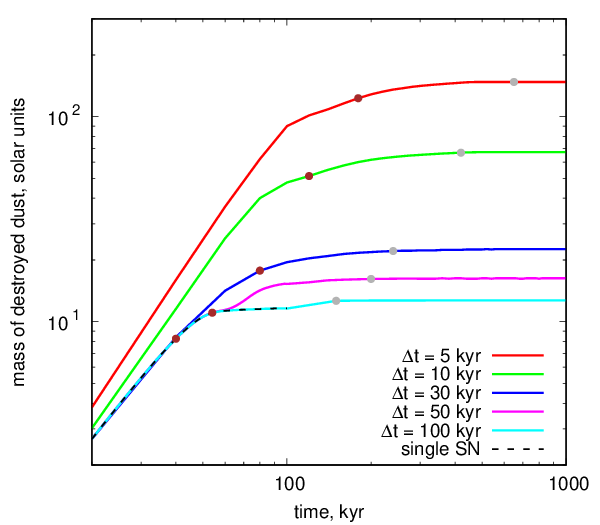}
\caption{
Mass of dust destroyed in the bubble for the models presented in Fig.~\ref{fig-mgas-evol}. The brown points correspond to the saturation time according to eq.~\ref{eq-sat}, whereas the grey points refer to the onset of radiative phase of bubbles ($\upsilon \sim 100$ km s$^{-1}$). The dashed line (only until 100~kyr) shows the evolution for the bubble formed by single SN explosion.
}
\label{fig-md-evol}
\end{figure}

The mass of interstellar dust destroyed behind the front evolves as in Fig.~\ref{fig-md-evol}. For low SN rate it follows the dependence for a single SN. For higher rates the destroyed dust mass increases linearly owing to dust grains entering behind the front get into hot post-shock gas with temperature of $\simgt 10^6$ K, where they are efficiently sputtered. Later, the rate of increase slows down and eventually saturates approximately as $\Delta t^{-1/2}$. Therefore, the flat part of the curves in Figure~\ref{fig-md-evol} represents the saturation regime when the total mass of destroyed dust does not increase while the SNe continue to explode. The saturation is linked to the fact that the shock speed falls below $\upsilon_0 \sim 200$ km s$^{-1}$, corresponding to post-shock temperature of $\sim 10^6$ K. Using the well-known wind solution for the bubble radius $r \propto t^{3/5}$ or for the expansion velocity $\upsilon \propto t^{-2/5}$ \citep{Castor1975}, the saturation in dust destruction occurs at a time scale, given by, 
\be 
t_{sat}\sim \upsilon_0^{-5/2}\left({\rho \Delta t\over E}\right)^{-1/2}\,.
\label{eq-sat} 
\ee 
The brown points in Fig.~\ref{fig-md-evol} correspond to this epoch, with shock velocity $\upsilon_0=200$~km s$^{-1}$. The curves show that saturation is reached at longer time scales for higher SN rate, i.e. in bubbles around more massive stellar clusters.  Incidentally, this also happens to be close to the epoch when a multiple SNe bubble of radius $R\sim (Et^3 / \Delta t \rho)^{1/5}$ becomes radiative  when the velocity is of the order of $v_0\sim 100$~km~s$^{-1}$ because at this point the post-shock temperature corresponds to a peak in the cooling curve.

In young adiabatic SN remnants, the dust is efficiently sputtered in the shell behind the shock front. Later, when the shell becomes radiative, the dust destruction rate decreases, and the incoming dust gets to compensate for earlier losses of dust mass. This fact manifests itself in a low value of the dust-to-gas ratio $\rho_d/\rho_g$ in the shell, when the shock velocity $\upsilon_s \simgt 200$~km/s, and which increases when the gas behind the shock cools down to $T_s\simlt 10^6$~K. In bubbles driven by multiple SNe, one has the same scenario. (see details in Appendix and in Fig~\ref{fig-d-pro}). 

As one can see, in the bubble formed by multiple SNe with rate $(\Delta t)^{-1} \sim (100~{\rm kyr})^{-1}$ the mass of destroyed dust reaches $\sim 13~\msun$ after explosions of three SNe at $t\sim 200$~kyr and does not grow at later time. Note that a single isolated SN destroys roughly the same mass of dust: $M_1 \sim 13~\msun$.  Using the established prescription \citep{McKee1989,Draine2009}, namely, if three SNe explode in isolation, one expects that these three SNe in total would destroy $\sim 3M_1 \sim 39~\msun$ of dust and  this value, therefore, increases as $N_{SN}M_1$. However, as one can conclude from Fig.~\ref{fig-md-evol}, the mass of destroyed dust saturates at some level and subsequent SNe do not destroy dust. For example, during the evolution of a cluster with $M_\ast \sim 4\times 10^4\msun$ ($\Delta t=100$~kyr) the total number of SNe is $N_{SN} \sim 250$. If each of these SNe had exploded in an isolated manner, they would have destroyed in total $N_{SN} M_1=3\times 10^3 \msun$. However, Fig.~\ref{fig-md-evol} shows that this mass is about $\sim 15~\msun$, which is about $200$ times lower than the expected value. 

Clusters with masses in the range $M_\ast \sim (4-12)\times 10^4\msun$ with a time delay $\Delta t \sim 30 - 100$~kyr and the mean number of SNe of $N_{SN} \sim 500$ per cluster would have destroyed in total $\sim 6.5\times 10^4\msun$  if they had exploded in an isolated manner. However, according to our calculations such a cluster destroys only $\sim 20\msun$ (it lies between cyan and blue lines in Fig.~\ref{fig-md-evol}).

\begin{figure}
\center
\includegraphics[width=8.5cm]{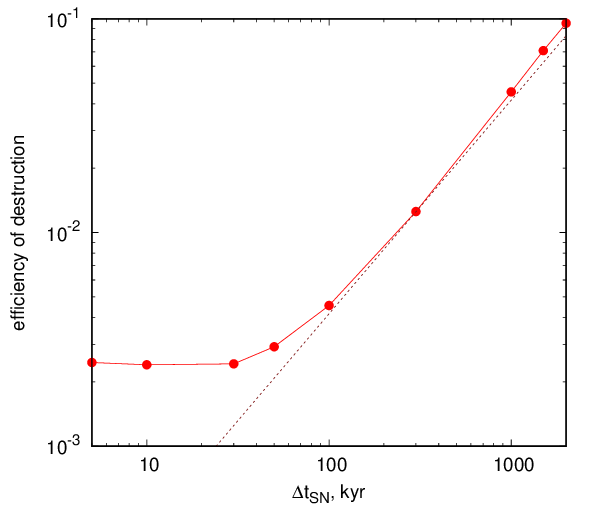}
\caption{
The relative dust destruction efficiency, i.e. the ratio of the the total mass of dust destroyed by SNe in cluster to the value of dust destroyed by the same number of isolated SNe. The dash line follows $\Delta t/t_{lf,max}$, where $t_{lf,max} \sim 24$~Myr.
}
\label{fig-ed-sat-dt}
\end{figure}

SNe exploding in clusters of $M_\ast \simlt 4\times 10^4\msun$ with a time delay $\Delta t \simgt 100$~kyr destroy the same mass of dust as that by a single SN: $M_1 \sim 13~\msun$. This remains valid unless the time delay between subsequent SNe is shorter than the fadeaway time (when the SN expansion velocity has slowed to the sound speed in the preshock medium $c_s$), which is $t_f \sim 1$~Myr for a single SN evolved in gas with $n\sim 1$~cm$^{-3}$ and $c_s \sim 10$~km/s for $T\sim 10^4$K \citep{Draine-book}. After $t_f \simgt 1$~Myr the SN cavity is filled by the gas from the SN shell. A shock from the next SN propagates again into the gas already processed by the previous SN shocks. During this process the interstellar dust swept-up in the shell can also partially enter the warm ($T\sim 10^4-10^5$~K) and low-density ($n\sim 0.1$~cm$^{-3}$) cavity. In our numerical calculations the mass of such dust reaches around $\simlt 1-2~\msun$ for the delay between SNe longer than $\sim 1$~Myr. This dust is destroyed by subsequent SNe. Thus, in such low-mass clusters the mass of the destroyed dust is close to that by a single SN, i.e. it differs slightly from $M_1$.

At longer time delay $\Delta t \sim 3$~Myr in very low-mass clusters of $M_\ast \sim 10^3\msun$, the cavity can be filled not only by the gas of the shell, but also presumably by the interstellar gas containing dust unprocessed by SN. This matter can be carried on into the cavity by interstellar turbulent flows. In our calculations we do not include such flows. However, one can estimate that for (super)sonic turbulence the cavity can be filled by the unprocessed interstellar material within a few million years. This issue will be studied in detail elsewhere. Therefore, a next SN explodes in the medium with properties (density, dust fraction) close to those being in the medium before the first SN has exploded. Thus, each SN in such low-mass clusters destroys the same dust mass as that by a set of SNe explode in isolation, it can be estimated as $N_{SN} M_1 \sim (M_\ast/150~\msun) M_1$.

The relative destruction efficiency can be determined as the ratio of the total mass of dust destroyed by SNe in cluster to the value of dust destroyed by the same number of isolated SNe $M_1 \times N_{SN}$. This efficiency is shown in Fig.~\ref{fig-ed-sat-dt} as a function of $\Delta t$. We find that for large $\Delta t$, or equivalently, small number of SNe, the efficiency of dust destruction increases, while for large number of SNe in clusters more massive than $M_\ast \simgt 10^5\msun$ with a time delay $\Delta t \simlt 30$~kyr, it reaches an asymptotic value of $\sim 0.0025$ (corresponding to the flat part of the solid line in Fig.~\ref{fig-ed-sat-dt}).

Another asymptotics is reached for intermediate and low-mass clusters of $M_\ast \simlt 4\times 10^4\msun$ with a time delay $\Delta t \simgt 100$~kyr (Fig.~\ref{fig-ed-sat-dt}). All SNe in a single such cluster are able to destroy the same mass of dust as an isolated SN. The destruction efficiency follows $\propto \Delta t/t_{lf,max}$, where $t_{lf,max} \sim 24$~Myr is the lifetime of the least massive star of $8\msun$, which can be a SN progenitor. A small difference of our numerical models for $\Delta t \simgt 1$~Myr from the simple linear function depicted by dash line comes from the destruction of the dust entered to the cavity during fadeaway phase. 

At very low-mass clusters of $M_\ast \sim 10^3\msun$ with a time delay a few times longer than the fadeaway time, i.e. $\Delta t \sim 3$~Myr, the cavity is assumed to be filled by the unprocessed interstellar material owing to the (super)sonic interstellar turbulence. Thus, we expect that each SN in such low-mass clusters destroys the dust mass estimated as $(M_\ast/150~\msun)M_1$.

Thus, in the bubbles formed around massive clusters of $M_\ast \simgt 4\times 10^4\msun$ or with average time delay between SNe explosions $\Delta t \simlt 100$~kyr, the efficiency of dust destruction is equal about $\epsilon_d \sim 0.004$. For lower mass clusters of $M_\ast \simlt 4\times 10^4\msun$ the efficiency is proportional to $\epsilon_d \propto \Delta t/t_{lf,max}$. Finally, for very low-mass clusters of $M_\ast \sim 10^3\msun$ it is expected to be close to unity.

The mass fraction of clusters exceeding $M_c \simgt 4\times 10^4\msun$, i.e. with delay $\Delta t \simlt 100~{\rm kyr}$, is
\be
 f_{h} \sim {\int_{M_c}^{M_{max}} M^{-\alpha+1} dM \over \int_{M_{min}}^{M_{max}} M^{-\alpha+1} dM } \approx 0.6\,, 
 \label{eqn:frac}
\ee 
assuming the limits $M_{min}$, $M_{max}$ to be equal to $10^3\msun$ and $3\times 10^7\msun$, respectively, and the index $\alpha\sim 2$ \citep{Krumholz2019}. Since the number of SN in each cluster with mass of $M_\ast$ is $N_{SN} \propto M_\ast$, thus, the fraction number of SNe in clusters massive more than $M_c$ is $f_{SN,h}\sim f_h \sim 0.6$.

Summarizing, the interstellar dust destroyed in the bubbles driven by multiple SNe in stellar clusters is estimated as $\sim \int \epsilon_d(M_\ast) N_{SN}(M_\ast) M_1 dM_\ast$, where $\epsilon_d$ is the destruction efficiency in a cluster with mass of $M_\ast$. In high mass clusters of $M_\ast \simgt 4\times 10^4\msun$ with $\Delta t \simlt 100$~kyr the mass of the destroyed dust can be estimated as $\sim \epsilon_{d,h} N_{SN,h} M_1$, where $\epsilon_{d,h} \sim 0.0025$ (see Fig.~\ref{fig-ed-sat-dt}), $N_{SN,h}$ is the total number of SNe in the high-mass clusters of $M_\ast \simgt 4\times 10^4\msun$, $N_{SN,h} \simeq f_{SN,h} N_{SN,tot}$, $N_{SN,tot}$ is the total number of SNe in all clusters, $f_{SN,h} \sim 0.6$ from eq.~\ref{eqn:frac}. Each cluster with $\Delta t$ from $\sim 100$~kyr to a few Myr destroys the dust mass equal to $M_1$. Thus, the mass of the destroyed dust is $\sim \epsilon_{d,m} N_{SN,m} M_1$, where $\epsilon_{d,m} \sim \Delta t/24~{\rm Myr}$ (see Fig.~\ref{fig-ed-sat-dt}), $N_{SN,m} \simeq f_{SN,m} N_{SN,tot}$, $f_{SN,m} \sim 0.3$ from eq.~\ref{eqn:frac} using the appropriate limits. SNe in clusters $M_\ast \sim (1-2)\times 10^3\msun$ is assumed to destroy the dust mass with $\epsilon_{d,l} \sim 1$, the fraction number of SNe in such clusters is $f_{SN,l} \sim 0.1$. 

Thus, the total mass of the interstellar dust destroyed in bubbles around clusters can be estimated as $(\epsilon_{d,h} f_{SN,h} + \epsilon_{d,m} f_{SN,m} + \epsilon_{d,l} f_{SN,l}) N_{SN,tot} M_1$. Note that the mass destroyed by the same number of SNe exploding in isolation is $\sim N_{SN,tot} M_1$. Thus, the ratio between the two is the efficiency of dust destruction, $\epsilon = \epsilon_{d,h} f_{SN,h} + \epsilon_{d,m} f_{SN,m} + \epsilon_{d,l} f_{SN,l}$. The first term is about $2.5\times 10^{-4}$, as one see below this value is negligible compared to the others. The second term increases linearly with $\Delta t$. We can estimate the second term for an average mass cluster of $M_\ast \sim 9\times 10^3~\msun$ with $\Delta t \sim 400$~kyr (the average value corresponds to the half of the fraction number of SNe in clusters of $M_\ast \sim 2\times 10^3 - 4\times 10^4\msun$), $\epsilon_{d,m} \sim 0.016$ (see also Fig.~\ref{fig-ed-sat-dt}), which gives the second term is $\sim 4.5\times10^{-3} \ll 0.1$. The third term for clusters $M_\ast \sim (1-2)\times 10^3\msun$ gives the major contribution, which is $\sim 0.1$ for $\epsilon_{d,l} \sim 1$. In case of lower efficiency value,  the ratio $\epsilon$ decreases. Therefore, the total ratio is $\sim 0.1$ for the cluster mass function with the index $\alpha\sim 2$ \citep{Krumholz2019}. Hence, the dust lifetime in the interstellar medium becomes longer by a factor $\sim 10$ compared to the value derived from the isolated SN evolution \citep{McKee1989}.

Thus, the total amount of dust produced by stars alone is sufficient to account for the interstellar dust budget, despite large amount of dust destroyed in SN shocks. While the dust destruction rate obtained both in models and observations \citep{McKee1989,Bianchi2007,Temim2015,Biscaro2016} are likely to be overestimated \citep[see also for review][]{Calura2025}. Indeed, our results indicate a lower destruction rate by taking into account the fact that most SNe occur in clusters and the effect of multiple SNe for dust destruction is different from that of an isolated SN.

Note that we have used equal intervals between SNe: $\Delta t \sim 24~{\rm Myr} / N_{SN}$. In massive clusters the number of stars being SN progenitors is large enough to produce frequent SNe explosions with more or less constant rate within several million years after the most massive star has exploded. At the same time in low-mass clusters there are a few massive stars, i.e. the initial mass function (IMF) has large gaps in the range $8-40~\msun$. Thus, the intervals between subsequent SNe explosions can be long enough and vary significantly from one stellar cluster to another. This manifests in the mass of dust destruction being dependent on the IMF of the cluster.

Although the efficiency of dust destruction is low in bubbles around massive clusters, their number is quite small compared to low-mass ones. Therefore, the destruction of the interstellar dust is controlled by SNe in low-mass clusters and the total budget of interstellar dust depends on the lowest mass $M_{min}$ of stellar cluster, where at least one star massive as $M>8\msun$ can be formed.     

\begin{figure}
\center
\includegraphics[width=8.5cm]{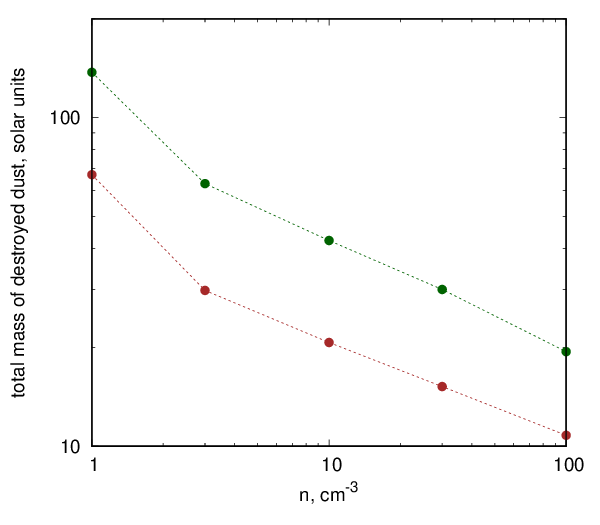}
\caption{
Total mass of dust destroyed in the bubble driven by SNe with the rate one per $\Delta t = 10$ and 5~kyr (green and brown lines, respectively) in the medium with gas number density $n$.
}
\label{fig-md-sat-n}
\end{figure}

The results mentioned so far pertains to an ambient density $n=1$ cm$^{-3}$. Next, we vary the ambient density in our simulation runs. Fig.~\ref{fig-md-sat-n} presents the total mass of dust destroyed in the bubble driven by SNe per $\Delta t = 10$ and 5~kyr as a function of the ambient gas density $n$. The curve shows that the destroyed dust mass decreases approximately as $M_d\propto n^{-1/3}$. In a denser environment, the shock slows down faster, and consequently, sputtering ceases at an earlier epoch, leading to less efficiency of dust destruction. Since the shock speed scales as $\upsilon \propto n^{-1/5} t^{-2/5}$, roughly, the epoch when sputtering is considerably diminished, i.e. when the wind bubble shock speed reaches a value $\upsilon_0\sim 200$~km/s, approximately scales as $t_{sp} \propto n^{-1/2}$. The mass of destroyed dust scales roughly as $\propto n R(t_{sp})^3$, or equivalently, as $\propto n^{-1/2}$ (accounting for the fact that the bubble radius reaches $R(t_{sp}) \propto (t_{sp}^3/n)^{1/5}$). This is  indeed the slope in Fig.~\ref{fig-md-sat-n} for $n \simlt 10$ cm$^{-3}$.

\subsection{Sequential SN explosions in a clumpy environment} 
\label{ssec:clumpy}

In the foregoing sections, we considered dust destruction in the bubble that evolve in a uniform medium. In a clumpy medium the bubble dynamics differs because the shock penetrates through a medium with finite-sized obstacles that are randomly distributed along its path. When a shock wave encounters a dense clump it drives a converging shock wave inwards with  velocity $\upsilon_{in}\sim \chi^{-1/2}\upsilon_{out}$ \citep{Bychkov1975}, where $\chi=\rho_{in}/\rho_{out}$ is the density contrast ({\it i.e.}, the ratio of the clump $\rho_{in}$ to the interclump $\rho_{out}$ density), and $\upsilon_{out}=\upsilon_s$ is the shock velocity. The parts of the shock that lie outside the projected clump area, flow around the clump and penetrate into the low-density interclump gas in the form of `long tongues' \citep[e.g.][]{Poludnenko2002,Korolev2015,Slavin2017,Wang2018}. As a result, the shock wave between clumps remains adiabatic on a longer time scale, and propagates through the diffuse low-density gas with a relatively higher velocity. In the self-consistent gas-dust dynamics the authors \citep{Dedikov2025na} have shown that the dust mass fraction destroyed in the bubble from a single SN in a clumpy medium is less than that in a homogeneous medium. This is because the dust confined in dense and partly-destroyed clumps does not suffer sputtering because the converging shock weakens as $\upsilon_s\propto \rho^{-1/2}$ \citep{Bychkov1975}. It is important to stress that each clump around the SN experiences the shock impact only once. In other words, once engulfed, the clump remains safe from being shocked again.   

Note that recently \citet{Scheffler2026} have reported about increasing destruction efficiency of the interstellar dust under the action of a SN remnant expanding in denser medium. However, their analysis has been constrained by short period of 10~kyr after the SN explosion for various average densities of the interstellar gas. At this age, dust is destroyed owing to the high expansion velocity of SN remnant. In reality, the destruction continues to the onset of the radiative phase in gas with density of $\sim 1$~cm$^{-3}$ \citep[see dash line in Fig.~\ref{fig-md-evol} and more details in][]{v2026}. Thus, \citet{Scheffler2026} likely underestimated the dust mass destroyed by SN evolving in a gas with average density of $\simlt 10$~cm$^{-3}$.
One might expect that at longer times they would end up increasing the efficiency of dust destruction by SN in a low-density interstellar medium and, in general, consistent with the results obtained in our work \citep[see the detailed discussion in][]{v2026}.

Although clumps around SN remnants experience shock impact only once, the clumps in bubbles driven by multiple SNe suffer impacts from  shocks episodically as long as the SNe explode. Each such encounter perturbs the medium enveloping the clump and strip it through the action of various hydrodynamic instabilities (Kelvin-Helmholtz, Rayleigh-Taylor and Richtmyer-Meshkov). This results in continuously repeated shock encounters, affecting the dust confined in clumps, and it increases the mass fraction of the destroyed dust. Despite this fact, the inner parts of clumps can survive continuous shock impacts due to radiative losses and strong density contrast. In such conditions, a substantial fraction of confined dust is protected against sputtering.

To study the dust destruction in multiple-SNe driven bubbles, we construct a clumpy gas density field by making use of the module pyFC\footnote{The code is available at https://bitbucket.org/pandante/pyfc/src/master/} \citep{Lewis2002} which generates lognormal `fractal cubes'. Within this model the gas density fluctuations in the ambient medium have a lognormal distribution and a Kolmogorov power-law spectrum with spatial index $\beta=5/3$ and maximum size of inhomogeneities equal to 16~pc. The density field is characterised by the mean $\langle \rho \rangle$ and the standard deviation $\sigma$ for logarithmic value. We consider a density field with $\sigma=2.2$, which corresponds to the contrast for 2$\sigma$ fluctuations exceeding a factor of $\simgt 80$ related to the mean value $\langle \rho \rangle = 1$~cm$^{-3}$. We assume pressure equilibrium at the initial state, i.e., $\rho T={\rm const}$, with zero gas and dust velocities. Thus, the gas temperature values differ by a factor of $\simgt 80$ between dense cloudlets and the intercloud medium.

\begin{figure*}
\center
\includegraphics[width=16cm]{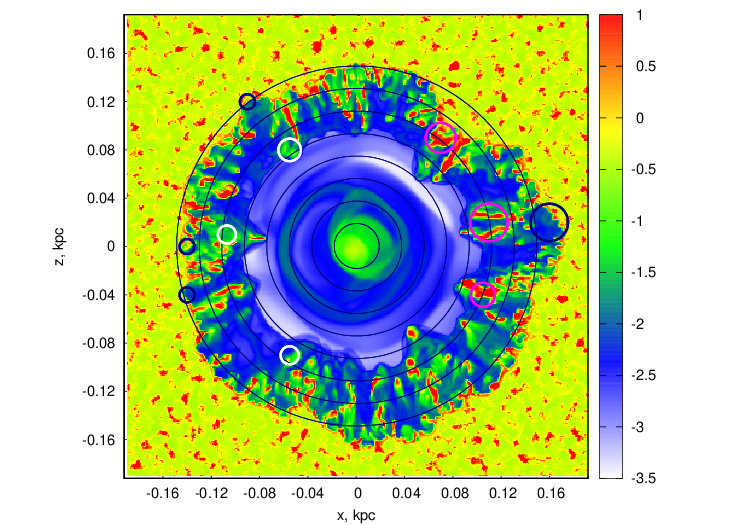}
\caption{
The $y=0$ plane slice of the gas density distribution in the bubble from clustered SNe {with the rate one per $\Delta t = 5$~kyr} in a clumpy medium with the average density $\langle\rho\rangle = 1~m_H$ g~cm$^{-3}$ at the age $t=800$~kyr. The colorbar shows the logarithm of gas density{, ${\rm log}(n,\ {\rm cm}^{-3})$}. Black circles roughly outline the forward shock at times $t=20,~60,~100,~200,~300,~440,~580$ and 800 kyr. Engulfed clumps being episodically impacted by upcoming shocks are seen to be deformed and compressed. At the same time, they are accelerated but because of compression they remain deep behind the forward shock. Small color circles are marked several examples of the regions with three density intervals: zones i are inside navy circles, zones ii are in magenta circles and zones iii are in white ones.
}
\label{fig-den-slice}
\end{figure*}

The gas behind the SNe shock front in a cloudy medium can be divided into three density intervals, which are connected with: i) the compressed low density gas immediately behind the irregularly shaped forward shock, ii) compressed gas of severely reshaped clouds deep behind the front, and iii) remnants of destroyed clouds that have penetrated into the  innermost parts of the bubble. Fig.~\ref{fig-den-slice} shows the gaseous regions connected with these three density intervals. The first is presented by a dense ($n>3$~cm$^{-3}$)  irregular shell. Several examples of these regions can be found inside the navy-colored circles in Fig.~\ref{fig-den-slice}. The next area adjoining and behind the shell consists of partly fragmented and in some cases merged dense ($n>10$~cm$^{-3}$) radially elongated clouds. One can see similar dense clumps in the magenta-colored circles. The third case is presented by low density ($n\simlt 0.1-1$~cm$^{-3}$) envelopes around the clouds that originate due to instabilities of their surface and episodic encounters of the shocks from multiple SNe, and by their remnants in the form of diluted fragments and filaments with density $n<0.03$~cm$^{-3}$. These fragments can be seen inside the white-colored regions in Fig.~\ref{fig-den-slice}.

The first region mostly consists of a gas locked initially in the diffuse interclump medium, and in those clumps that have been initially located  closer to the central area of the bubble and have been completely destroyed by the current epoch. The other two represent the material of the clouds that have been recently encountered by the forward shock and which have partly survived such an impact. Dust particles in these three regions experience different environmental influence, and in order to evaluate the fraction of surviving dust we have to approach them separately.  

At early periods ($t\simlt 200$~kyr), clumps located close to the innermost region are efficiently disintegrated with only a small part being swept up along with the expanding shell. The dust confined in such clouds is mostly destroyed.  At later epochs, when the forward shock decelerates below $30\hbox{--} 100$~km~$s^{-1}$ ($\Delta t = 100$ and $5$~kyr, respectively), the dense internal regions of the clouds survive along with the dust locked in them. A certain fraction of dust locked initially in clouds (the cloud dust) can leave their original reservoir and fill the hot intercloud medium, where the dust experiences strong sputtering. The cloud dust particles that remain in the low density and high temperature envelope of clouds are also expected to be destroyed.

\begin{figure}
\center
\includegraphics[width=8.5cm]{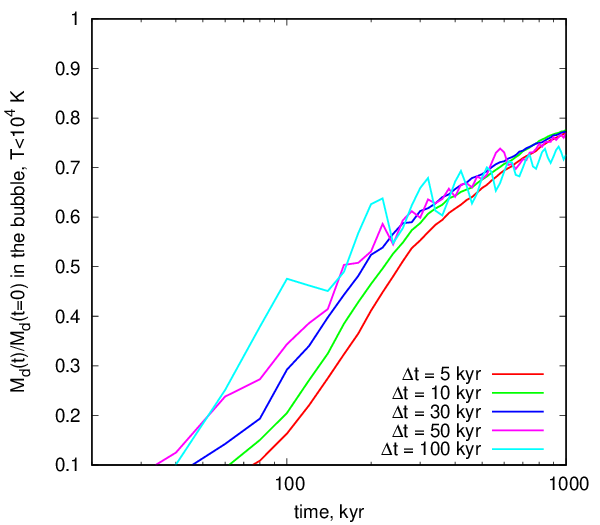}
\caption{
Mass fraction of the dust that can survive against sputtering, being locked {\it initially} (at $t=0$) in dense clouds ($\rho > \langle \rho \rangle$) inside the region occupied by the bubble till the epoch $t$, i.e. the ratio of the mass of dust {\it remained in the clouds} at time $t$ within the region outlined by the forward shock, $M_d(t)$, to the mass of dust locked in the clouds at $t=0$, $M_d(t=0)$. The bubble from multiple SNe explosions expands in a clumpy medium with an average density $\langle\rho\rangle = 1~m_H$ g~cm$^{-3}$. 
}
\label{fig-frac-md-t4}
\end{figure}

In order to understand the overall picture of dust destruction within the region occupied by the remnant produced by multiple SNe in a cloudy medium, we illustrate in Fig.~\ref{fig-frac-md-t4} how the dust locked initially (at $t=0$)\footnote{Among other parameters each superparticle has a unique id-number, therefore, we can trace the path of each particle.} in the dense clouds is destroyed inside the region occupied by the bubble at the epoch $t$. More explicitly, this is the ratio of the mass of  dust {\it that remain in the clouds} at time $t$ within the region outlined by the forward shock, $M_d(t)$, to the mass of dust ($M_d(t=0)$) locked in the clouds at $t=0$. 
Note that $M_d(t)$ {\it does not account those dust particles that escaped} the original clouds and spread through the intercloud medium $\rho<\langle\rho\rangle$. 

At early epochs $t<100$~kyr, the fraction of surviving dust is very small because the dense cloud gas is strongly heated and disintegrated by the shock, as can be observed in Fig.~\ref{fig-den-slice}. Correspondingly, dust associated with this gas does not survive much.
At later times, clumps partly survive disruption from shock waves, resulting in the growth of the surviving dust mass. Before $\sim 200$~kyr the fraction is greater for higher SNe rate. However, this dependence becomes negligible at later epochs. 

\begin{figure}
\center
\includegraphics[width=8.5cm]{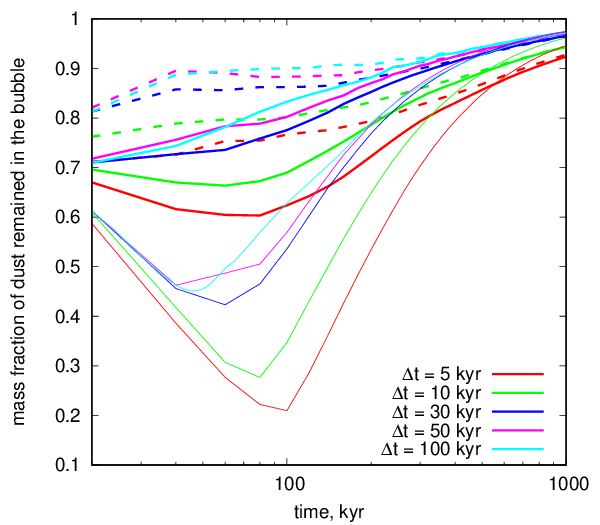}
\caption{
Mass fraction of the dust surviving against sputtering in a gas inside {\it the entire bubble outlined by the forward shock} driven by clustered SNe in a clumpy medium with the average density $\langle\rho\rangle = 1~m_H$ g~cm$^{-3}$ (thick lines) and in a homogeneous medium with the same density (thin lines). Thick dashed lines depict the mass fraction of the dust inside clumps with density three times higher than the average value.
}
\label{fig-frac-md}
\end{figure}

Another characteristic that describes the dust processing behind shocks is the mass {\it fraction of all dust} that survive against sputtering inside {\it the entire bubble outlined by the forward shock}. Fig.~\ref{fig-frac-md} presents this dependence by thick solid lines. At early times $t\sim 70$~kyr, this fraction decreases for the case of one SN per $\Delta t=5$~kyr (red line). Later it begins to grow, thanks to dense lumps that survive in the bubble (Fig.~\ref{fig-frac-md-t4}). In a clumpy medium, a significant mass of dust contains in dense fragments, which are efficiently stripped only at early epochs. The mass fraction of the dust surviving against sputtering and associated with such dense clumps increases (see thick dashed lines in Fig.~\ref{fig-frac-md}). Such gas includes dense lumps, compressed fragments and dense irregular shell. Later, the mass-loading of the flow behind the forward shock increases and the major part of the surviving dust remains associated with gas in dense clumps. This is shown by converging of thick dashed and solid lines in Fig.~\ref{fig-frac-md} at later times. For lower SN rate, the picture is qualitatively similar, however, more dust survives in the bubble because the destruction of dense fragments is less effective.

In  the bubble expanding in a homogeneous medium, dust is efficiently destroyed by thermal and kinetic sputtering in the bubble's shell, while the the forward shock remains adiabatic. When the shock enters the radiative stage, dust particles decelerate in the dense cold layers due to drag force, and where they become protected against sputtering \citep{vs2024}.

When the bubble expands in a cloudy medium dust is destroyed (a) in completely destroyed clumps by the forward shock, (b) in the shell formed immediately behind the forward shock and (c) in the hot envelopes around lumps far behind the front due to stripping by shocks from continuous SNe explosions. The first two mechanisms play a role when the forward shock is quite strong and with the velocity higher than 200~km/s. In a clumpy medium the forward shock moves presumably in the interclump low-dense medium and maintains high velocity longer. 
Those parts of the shock that interact with the clumps and penetrate them with lower velocity and in which the dust locked in the denser clumps, remain protected against sputtering. Continuous SNe explosions result in multiple shocks so that high-velocity sound waves impinge on dense clumps that have survived the impact of the forward shock. These waves strip the clumps and form hot envelopes around them. Thus, deep behind the forward shock dust continues to income from destroyed gaseous fragments. As seen in Fig.~\ref{fig-frac-md}, the first two processes (a) and (b) operate during about 300~kyr. The third one comes into play when all the remaining dust becomes locked in dense gas.

During the first several hundreds of kyrs the fraction of surviving dust $f_{surv}$ remains higher in the bubble expanding in a clumpy medium than in the one expanding in a homogeneous one (Fig.~\ref{fig-frac-md}). Later on, $f_{surv}$ in clumpy environment decreases because of  continuous stripping of dense fragments (as seen in Fig.~\ref{fig-frac-md}). Eventually at times $t>100$~kyr the rate of surviving dust in a homogeneous medium becomes higher, because of the process (c) (of dust getting locked in dense clumps and destruction occurring only in the envelopes around them) becoming dominant at this phase, and at $t \simgt 300$~kyr the surviving fraction of dust in both cases turns to become of the same order.   

\begin{figure}
\center
\includegraphics[width=8.5cm]{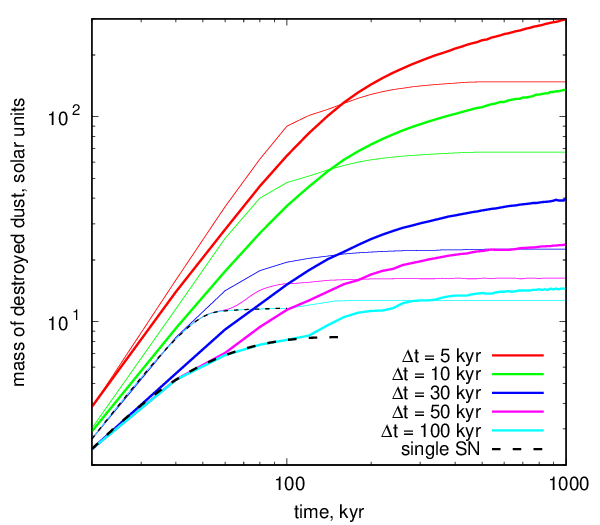}
\caption{
Mass of dust destroyed in the bubble evolved in a clumpy medium (thick lines) with mean density $\langle \rho \rangle = 1$~cm$^{-3}$, the standard deviation $\sigma=2.2$ for logarithmic value and maximum size of inhomogeneities equal to 16~pc. Thin lines depict the dust mass evolution in the bubble expanded in a homogeneous medium presented in Fig.~\ref{fig-md-evol}.
}
\label{fig-md-evol-cloudy}
\end{figure}

Fig.~\ref{fig-md-evol-cloudy} presents the mass of dust destroyed in the bubble for the same SN rates as shown in  Fig.~\ref{fig-md-evol}, but evolving in a clumpy medium. At early times, this mass (of destroyed dust) in the bubble evolving in a clumpy medium is lower than that in a homogeneous one. For SN rate $(\Delta t)^{-1} \sim (100~{\rm kyr})^{-1}$ the mass of dust destroyed during expansion in a clumpy medium begins to exceed the value reached in a homogeneous medium at an age of the bubble $\sim 300$~kyr. For higher SN rate this crossing occurs earlier, at $\approx 150$~kyr, for all other rates.

\begin{figure}
\center
\includegraphics[width=8.5cm]{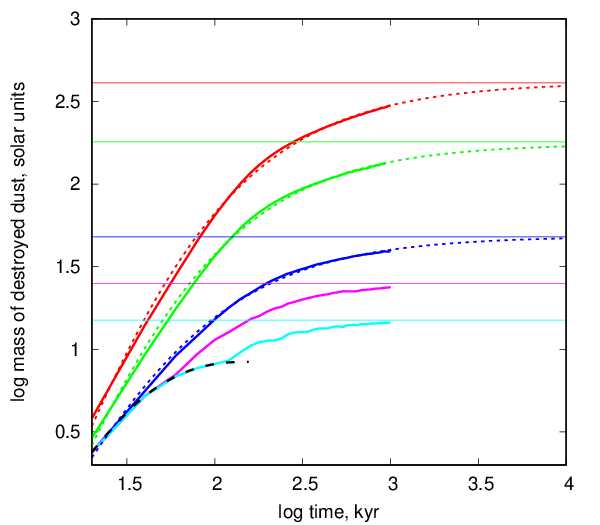}
\caption{
Mass of dust destroyed in the bubble evolved in a clumpy medium depicted in Fig.~\ref{fig-md-evol-cloudy} (thick lines) and their fits (dotted lines, see the text for details). Thin lines show the saturation limits. The dashed line shows the mass of dust destroyed by a single SN.
}
\label{fig-md-evol-cloudy-appro}
\end{figure}

We have earlier seen that in a homogeneous medium the mass of dust destroyed in the bubble becomes saturated after the shock speed falls below 200~km/s (see eq.~\ref{eq-sat}). In a clumpy medium, dense fragments located deep behind the forward shock can supply dust into hot gas even after the forward shock decelerates below 200~km/s. Such fragments are stripped by shocks from ongoing SN explosions in the bubble. The characteristic time of this process depends on the density contrast roughly as $\propto \chi^{1/2}$. When outer layers of clumps stripped, the density contrast can reach $\simgt 10^2-10^3$ (Fig.~\ref{fig-den-slice}). Further development of the Kelvin-Helmholtz instability becomes suppressed and the destruction time for lumps of size $\sim 10$~pc (Fig.~\ref{fig-den-slice}) exceeds tens of Myr \citep[e.g.][]{Klein1994,Poludnenko2002}. Consequently, the dust mass destroyed in the bubble in a cloudy medium also saturates as in the homogeneous scenario, though at later ages. An increase in the SN rate leads to saturation at a later epoch. For instance, for the rate one SN per $\Delta t=100$~kyr this has already reached till the age of $\sim 800$~kyr, and for $\Delta t=50$~kyr this is expected to be at $\sim 1-1.5$~Myr. We anticipate similar behaviour for higher SN rate, which allows us to fit further evolution of the mass as $f(x) = a/[1+{\rm exp}(-b(x-c))] - 0.5 a$, where $x = {\rm log} (t, {\rm kyr})$, $f(x) = {\rm log} (M_d, \msun)$, $a,\ b,\ c$ are parameters. Fig.~\ref{fig-md-evol-cloudy-appro} presents the fits for three high values of SN rate. The fits come to saturation at $\sim 10$~Myr, which is several times longer than that in a homogeneous medium.

\begin{figure}
\center
\includegraphics[width=8.5cm]{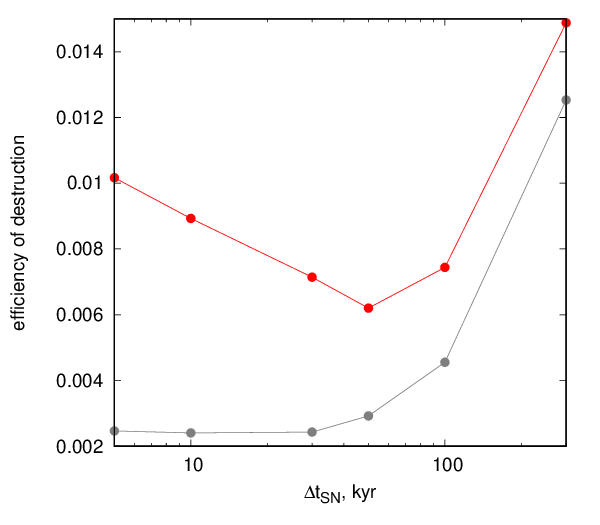}
\caption{
The dust destruction efficiency for the bubble evolved in a clumpy medium (red line) with mean density $\langle \rho \rangle = 1$~cm$^{-3}$, the standard deviation $\sigma=2.2$ for logarithmic value and maximum size of inhomogeneities equal to 16~pc. Grey line depicts the efficiency for the bubble evolved in a homogeneous medium as shown in Fig.~\ref{fig-ed-sat-dt}.
}
\label{fig-ed-sat-dt-cloudy}
\end{figure}

Accounting for the larger mass of dust destroyed in the bubble in a clumpy medium (Fig.~\ref{fig-md-evol-cloudy}), the dust destruction efficiency increases and becomes about $1.5\hbox{--}4$ times higher in a clumpy medium compared to that in a homogeneous medium (compare red and grey lines in Fig.~\ref{fig-ed-sat-dt-cloudy}). The efficiency is increased for higher SN rate due to more frequent stripping events by impacts of clumps with high-velocity waves coming from SNe explosions. Smaller inhomogeneity in density (smaller value of density dispersion $\sigma$) leads to a decrease in this efficiency and shortening of the period for reaching saturation. For lower SNe rate the efficiency is expected to follow $\Delta t/t_{lf,max}$ similar to that in a homogeneous medium (Fig.~\ref{fig-ed-sat-dt}).

\section{Discussion} \label{sec:discus}

Above we have considered the destruction of the interstellar dust in multiple SN-driven bubbles evolved in an unstratified medium. However, stellar clusters are usually formed in relatively thin galactic disks, in which the gas density decreases with height over the disk midplane. It is well known that in these condition, the expansion of bubbles in a vertically stratified medium is affected by the decrease in density, and bubbles can break out of the disk when its radius reaches the scale height \citep[e.g.,][]{Kovalenko1985,McCray1987,Fielding2018}. This results in forming HI holes, which are found in large number in both dwarf and face-on spiral galaxies \citep[e.g.,][etc]{Heiles1979,Kamphuis1991,Walter1999,Bagetakos2011,Egorov2014,Egorov2017,Robertson2025,Shaji2025}. Even though in most cases described in this paper, the asymptotic behaviour of dust evolution is reached at shell radii below the typical scale height for galactic interstellar medium, here we will discuss qualitatively possible effects of vertical stratification on aspects relevant here.

In the vertical direction the shell of the bubble is thin and and moves upward with increasing velocity $v_z\propto {\rm exp}(2z/3z_0)$. The shell inolves a relatively small fraction of the bubble. In contrast, those parts of the shell expanding predominantly in the disk are thicker and more massive with degrading velocity \citep[e.g.,][]{vs2022,Fielding2018}. Dust grains pre-existing in the interstellar meduim are swept up in the bubble shell. Due to their inertia they enter into hot gas far behind the shock, where they are efficiently sputtered \citep{vs2024,Dedikov2025irx,Dedikov2025na}. In the vertical direction, grains might penetrate deeper toward the bubble interior because the shell in this part remains adiabatic and moves with higher velocity \citep[e.g.,][]{vs2022}. In this case, two possible effects can come into play. On one hand, since dust density decrease along with gas in a stratified ISM, the attenuation in the upper part of the HI hole is lower than in a non-stratified ISM. This effect becomes more remarkable in HI holes around massive stellar clusters having mass greater than several $10^4\msun$ and/or evolving in thin disks. On the other hand, since the gas temperature increases in the vertical direction as $\sim {\rm exp}(4z/3z_0)$, dust particles in the upper parts of the shell are hotter, with temperature $T_d\sim {\rm exp}(2z/9z_0)$. This effect can, in principle, be distinguished in huge HI holes  observed in face-on galaxies \citep[e.g.,][]{Kamphuis1991} and in overlapping galaxies \citep[e.g.,][]{Keel2001,Robertson2025}. However, more detailed numerical analysis needed for a precise conclusion.

Possible variations of dust properties inside/outside and between HI holes can charactirize the destruction efficiency of grains by shocks driven by (super)bubbles. Therefore, dust maps obtained in backlit galaxies might be used to help us for studying the morphology of such variations \citep{Keel2023,Robertson2025}. However, such large structures can be used for understanding possible disk break-out in more details, especially in extended galactic disks \citep[e.g.,][]{Holwerda2009}. Stronger sputtering of dust grains in upper parts of HI holes in a stratified ISM results in surviving of large grains only, that can be revealed in changing reddening of starlight from the underlying stellar population. Spatial variation of the $R_V$ found by \citep{Fitzpatrick1990,Fitzpatrick1999met,Fitzpatrick1999,Fitzpatrick2004} can be used to scrutinize dust properties in the internal and peripheral parts of HI holes, and to understand the effects of vertical stratification on dust properties. Synthetic $A_V$ and $R_V$ maps can be obtained from numerical simulations of the 3D gas+dust evolution of the multiple SNe-driven bubbles in a stratified medium similar to those shown in Fig.~\ref{fig-den-slice}. Such models will be carried out elsewhere. 

It is well established that SN remnants are sources of IR emission \citep[e.g.][]{Dwek1987,Graham1987,Arendt1989,Meixner2006,Seok2013,Seok2015,Chawner2019,Chawner2020}. In young SN remnants such emission comes from dust produced by SN \citep{Kozasa1989,Todini2001,Nozawa2006}. In older SN remnants the swept-up  interstellar dust is expected to contribute mainly to this emission \citep[e.g.][]{Dedikov2025irx}. SN remnants are bright IR sources until dust remains dispersed in hot gas inside the remnant. Dust grains located in hot gas can be destroyed by sputtering or move to gas in other thermal phases due to gas cooling or dynamical effects. Similar mechanisms occur in bubbles driven by multiple SNe explosions.

In the previous sections we explored the questions as to how interstellar grains are destroyed in SNe-driven bubbles and how destruction of the swept-up dust exhausts after the shell cools down. This dust passes rapidly across the hot layer and accumulates mostly in the cold shell, with a negligible amount left in the hot inner region. Thus, the swept-up dust does not contribute in the IR emission of the bubble after the dust stops to be destroyed in the bubble shell or at $t\simgt t_{sat}$ (eq.~\ref{eq-sat}), when the shell velocity is of the order of $v_0\sim 100$~km~s$^{-1}$. 

Meanwhile, dust is injected in hot interior of the bubble by every subsequent SN explosion. The injected grains can become a significant source of IR emission as long as they survive in this hostile environment. The grain lifetime is roughly $\tau \sim 10^5 a_{\rm \mu m} n_{\rm H}^{-1}$~yr, and if this timescale is shorter than the time between subsequent SN explosions $\Delta t_{SN}$, then dust cannot accumulate in the hot interior. However, dust particles can leave the hot gas due to their high velocity, grains are caught up with the bubble shell and move in to colder gas, where they cannot emit efficiently.

In the bubbles evolving in a clumpy medium, the dust mass in hot gas is replenished by not only swept-up unprocessed interstellar dust, but also from stripped dense gaseous lumps far behind the forward shock of the bubble. Owing to the latter process, dust can be dispersed in the hot gas on longer timescales,  exceeding several Myr, for SNe explosions more frequent than one SN per 30~kyr (see Fig.~\ref{fig-md-evol-cloudy-appro}). Therefore,  a bubble evolving in a clumpy medium can remain a bright IR source up to several Myr. Simultaneously, the hot plasma inside the bubble emits in X-ray. Due to continuous SNe explosions the plasma temperature is supported at $T_X \sim (2-4)\times 10^6$~K for  SNe rate of one SN per $\sim 100-5$~kyr in  bubbles older than several tens of kyr. Using the approach described in the paper \citep{Dedikov2025irx}, we calculate the ratio between the IR and X-ray luminosities averaged over the whole bubble driven by multi-SNe. Usually the IR-to-X-ray ratio is used to determine the contribution of dust cooling into radiative losses in hot plasma \citep{Dwek1987}. We find that the ratio typically ranges between $\sim 1\hbox{--}30$. These values weakly depend on the SNe rate. However, a decrease of density contrast (density dispersion $\sigma$) in the ambient medium results in more efficient cloud destruction and, therefore, diminishing IR emission down to almost negligible values, for bubble older 100~kyr evolved in a homogeneous medium with $n\sim 1$~cm$^{-3}$.

\section{Conclusion} \label{sec:concl}

We have considered the destruction of dust in shells driven by multiple SNe explosions in stellar clusters. Using 3D hydrodynamic simulations, we have investigated how the single-size interstellar dust grains are destroyed in the multi-SNe bubble expanding in a homogeneous as well as  clumpy medium. Our results can be summarized as follows:

\begin{itemize}
\item The mass of dust destroyed in the multi-SNe bubble does not grow after a certain epoch and becomes saturated, while SNe continue to explode. The timescale for this saturation is proportional to SNe rate in cluster;

\item The mass of dust destroyed in bubble per SN decreases for higher SN rate; the destruction efficiency -- the ratio of the the total mass of dust destroyed by clustered SNe to the value of dust destroyed by the same number of isolated SNe -- in bubbles evolving in a homogeneous medium drops for higher mass clusters. E.g., around clusters with $M_\ast \simgt 4\times 10^4\msun$, it is less 0.4\%. For lower mass clusters the efficiency is proportional to $\Delta t/t_{lf,max}$. Therefore, the destruction of the interstellar dust is controlled by SNe in low-mass clusters.

\item The mass of dust destroyed by multiple SNe in clusters approximately $10\%$ of that in an ensemble of equal number of isolated SN. Thus, the lifetime of dust in the ISM is longer by a factor $\sim 10$ compared to the value previously derived in the literature for isolated SN.

\item In a clumpy medium the efficiency increases by about a factor of $1.5$ in bubbles formed around clusters with $M_\ast \sim 4\times 10^4\msun$ and up to 4 times around $M_\ast \sim 8\times 10^5\msun$.
\end{itemize}


\section{Acknowledgements} \label{sec:ackn}
We thank the referee for the valuable comments.
We acknowledge Nina Sartorio and Florian Kirchschlager for very detailed and constructive comments which helped us to improve the presentation of our results.
We thank Yuri A. Shchekinov for many valuable comments, Svyatoslav Dedikov for discussions.

\appendix

\section{Radial profiles}

\begin{figure}
\center
\includegraphics[width=9cm]{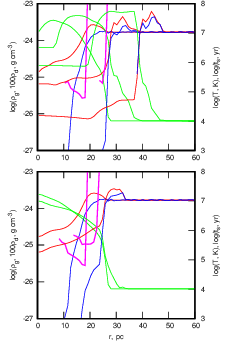}
\caption{
The angle-averaged radial profiles of gas density ($\rho_g$, red line), gas temperature ($T_g$, green line),  density of the interstellar dust ($\rho_d$ multiplied by a factor of 100, blue line) and sputtering timescale (magenta line) for the bubble formed by multiple SNe with rate $(\Delta t)^{-1} \sim (3\times 10^4~{\rm yr})^{-1}$ (upper panel) at bubble age $t=40$, 100 and 200~kyr (lines from left to right). The sputtering timescale at $t=200$~kyr is longer than $10^9$yr. In lower panel the profiles are shown for a single SN of age $t=40$, 100~kyr.
}
\label{fig-d-pro}
\end{figure}

Figure~\ref{fig-d-pro} presents the angle-averaged radial profiles of gas density, temperature, dust density and sputtering timescale for a bubble formed by multiple SNe with time delay $\Delta t \sim 30~{\rm kyr}$ (upper panel). Due to their inertia,  interstellar dust particles swept up together with gas by shock front move further into hot interior of the bubble \citep{vs2024}. There grains are subjected to sputtering. A supply of dust mass behind the front due to the bubble expansion is not sufficient to replenish its decrease by sputtering. This reveals in lower dust density in the shell at $t=40$~kyr: the dust density (blue lines) does not follow the gas density (red lines). It is clearly seen that the dust density decreases by about a half order of magnitude at the maximum gas density compared to that expected for the absence dust destruction.  If there had been no destruction of dust, then the densities of gas and dust would have followed each other (the red lines would coincide the blue ones).

When the shock velocity $v_s$ falls below $v_s < 200$~km~s$^{-1}$, i.e. the temperature $T_s < 10^6$K, the sputtering becomes weaker $\rho_d/\dot \rho_d>t$, the delay between dust and gas profiles diminishes and the ratio $\rho_d/\rho_g$ approaches a constant. Similar picture can be also found at the profiles for the bubble formed by a single SN explosion (lower panel).

\bibliographystyle{cas-model2-names}
\bibliography{p-bib1}

\end{document}